\documentclass[aps,prd,showpacs,twocolumn,groupedaddress,floatfix,nofootinbib]
{revtex4}

\usepackage{amsmath,amssymb}
\usepackage{textcomp}
\usepackage{hyperref}
\usepackage{bm}
\usepackage{graphicx}
\usepackage{psfrag}
\usepackage[usenames]{color}
\usepackage{empheq}
\usepackage{ulem}
\allowdisplaybreaks[4]

\newcommand{\uvec}[1]{\bm{\hat{#1}}}
\newcommand{\dvec}[1]{\dot{\bm{#1}}}

\newcommand{\duvec}[1]{\dot{\bm{\hat{#1}}}}

\newcommand{\be}{\begin{equation}}
\newcommand{\ee}{\end{equation}}

\newcommand{\up}{{\mbox{\tiny up}}}

\newcommand{\sub}[1]{_{\text{#1}}}
\newcommand{\tk}{t_{\bm{k}}}

\begin{document}



\title{Fast frequency-domain waveforms for spin-precessing binary inspirals}

\author{Antoine Klein}
\email{aklein@olemiss.edu}
\affiliation{Department of Physics and Astronomy, The University of 
Mississippi, 
University, Mississippi 38677, USA}
\affiliation{Department of Physics, Montana State University, Bozeman,
Montana 59717, USA}

\author{Neil Cornish}
\affiliation{Department of Physics, Montana State University, Bozeman,
Montana 59717, USA}

\author{Nicol\'as Yunes}
\affiliation{Department of Physics, Montana State University, Bozeman,
Montana 59717, USA}

\date{\today}

\begin{abstract}

The detection and characterization of gravitational wave signals from compact 
binary coalescence events relies on
accurate waveform templates in the frequency domain.
The stationary phase approximation (SPA) can be used to compute closed-form 
frequency-domain waveforms for nonprecessing, quasicircular
binary inspirals. However, until now, no fast frequency-domain waveforms have 
existed for generic, spin-precessing quasicircular
compact binary inspirals. Templates for these systems have had to be computed 
via a discrete Fourier transform of finely sampled
time-domain waveforms, which is far more computationally expensive than those 
constructed directly in the frequency domain, especially for those systems that 
are dominated by the inspiral part.
There are two obstacles to deriving frequency-domain waveforms for precessing 
systems: (i)
the spin-precession equations do not admit closed-form solutions for generic 
systems; (ii) the SPA fails catastrophically, i.e. in some situations, 
the second derivative of the signal phase vanishes, so that the direct 
application of the SPA predicts a divergent amplitude.
Presently there is no general solution to the first problem, so we must resort 
to numerical integration of the spin-orbit precession equations.
This is not a significant obstacle, as numerical integration on the slow 
precession timescale adds very little to the computational cost of generating 
the waveforms.
Our main result is to solve the second problem, by providing an alternative to 
the SPA that we call the method
of shifted uniform asymptotics (SUA) that cures the divergences in the SPA.
The construction of frequency-domain templates using the SUA can be orders of 
magnitude 
more efficient than the time-domain ones obtained through a discrete Fourier 
transform.
Moreover, this method is very faithful to the discrete Fourier 
transform, with mismatches on the order of $10^{-5}$. 

\end{abstract}


\maketitle
\section{Introduction}

In the two-body problem of 
general relativity, the coupling of the spin and orbital angular momenta leads 
to Lense-Thirring precession~\cite{poissonwill}, which causes the 
orbital plane as well 
as the spins of the bodies to change orientation. In the Solar System, such an 
effect  is often negligible, since the magnitude of the spin angular momenta is 
usually much smaller than that of the orbital angular momentum.
However in the late inspiral of compact objects, such as black holes and 
neutron stars, 
this is not necessarily the case, and the precession of the orbital plane can 
lead to very complicated motion.  

Precession is both a blessing and a curse. It adds rich structure to the 
templates which can greatly enhance
the accuracy of parameter estimation~\cite{vecchio:2004,Lang:1900bz,stavridis,
Yagi:2009zm, Klein:2009gza,Lang:2011je}, but it 
also makes detecting and 
characterizing the signals more
challenging.  From a practical standpoint, part of the challenge is that it is 
not easy to produce accurate
waveform templates for precessing systems at reasonable computational cost.
These waveform templates require precise knowledge of the 
evolution of the orbital angular momentum, the orbital phase, and the 
frequency. 
The fact  that no analytic solution to the precession equations is known in 
full 
generality has been used to justify the use of approximate solutions, which may 
suffice
for detection purposes, but are inadequate for parameter estimation.

The importance of precession in gravitational wave (GW) parameter estimation 
has been 
recognized early on in the development of accurate templates for data 
analysis~\cite{Cutler:1992tc,Cutler:1994ys,Apostolatos:1994mx,vecchio:2004,
Lang:1900bz,stavridis,
Yagi:2009zm, Klein:2009gza,Lang:2011je,Klein:2013}. Precession not only affects 
the evolution of the GW 
phase,  but it also introduces modulations to the GW amplitude that evolve on 
much longer time scales than
the orbital period. Such modulations have been shown to 
break parameter degeneracies that are unavoidable in 
nonprecessing templates~\cite{vecchio:2004, Lang:1900bz, 
Chatziioannou:2014bma,Chatziioannou:2014coa}. 
Physically, templates that can accurately reproduce the complicated structure 
of signals will in general be capable of extracting much more useful 
information 
than templates that do not capture this structure.  

Historically, the construction and study of precessing templates dates back to 
the beginning of the 21st century. 
Vecchio in 2004~\cite{vecchio:2004} showed that parameter estimation in the 
context of LISA could be greatly improved when using the 
analytical simple precession model derived by Apostolatos \emph{et 
al}.~\cite{Apostolatos:1994mx}. Later in 2006, Lang and 
Hughes~\cite{Lang:1900bz} proposed a prescription for waveform templates with 
generic precessing spins, which was later refined to include subdominant 
harmonics~\cite{Klein:2009gza} and then shown to improve parameter 
estimation when precession effects are moderately 
suppressed~\cite{Lang:2011je}. 
The inclusion of precession has recently been shown to improve parameter 
estimation 
for neutron star inspiral 
sources~\cite{Chatziioannou:2013,Chatziioannou:2014bma,Chatziioannou:2014coa}, 
as well as to strengthen tests of general relativity with 
GWs~\cite{stavridis,Yagi:2009zm}.

Despite these efforts to construct closed-form approximate 
templates,  current parameter estimation algorithms for advanced, ground-based 
interferometer data, 
such as advanced 
LIGO (aLIGO)~\cite{aLIGO} and advanced Virgo 
(AdV)~\cite{AdV}, use numerical time-domain templates 
for precessing systems, i.e.~templates constructed 
from the discrete Fourier transform (DFT) of time-domain, 
numerical solutions to the post-Newtonian (PN) expanded 
(and sometimes resummed) Einstein equations. In the PN approximation, the field 
equations are expanded in small velocities and weak fields and then 
possibly resummed~\cite{blanchet-review}. The construction 
of such templates is quite computationally expensive because, when computing
the DFT of a time series, one needs to discretize the signal with a uniform 
spacing that is capable of resolving the shortest
time scales. This fixes the time resolution to one 
half of the inverse of the highest frequency reached by the signal, which is 
typically much higher 
than the resolution needed to resolve the signal during most of the inspiral. 

Because of this computational cost issue, a significant amount of effort has 
recently been put into deriving precessing templates constructed directly in 
the frequency domain, i.e.~without having to compute the DFT of a time series,  
but that remain as accurate as 
possible~\cite{Apostolatos:1994mx,apostolatos1995,Lundgren:2013jla,Klein:2013,
Chatziioannou:2013}. Such 
efforts relied either on approximate solutions to the precession 
equations, or on solutions to simplified precession equations. In 
particular, frequency-domain waveforms were constructed 
for binaries with one spinning component~\cite{Lundgren:2013jla}, and 
for 
double
precessing binaries but in the small misalignment 
approximation~\cite{Klein:2013} and in the small spins 
approximation~\cite{Chatziioannou:2013}.
Templates for generic precessing 
binaries have also been constructed using single-spin 
formulae~\cite{Hannam:2013}, but this cannot reproduce every feature 
present in generic precessing systems which exhibit more complex 
evolution.

Once a closed-form prescription for the temporal evolution of the angular 
momentum is specified, the Fourier transform of the GW response function is 
conventionally modeled through the stationary phase approximation 
(SPA). This approximation assumes the Fourier transform at a given frequency 
is dominated by the Fourier integral in a small neighborhood 
about a certain stationary point, i.e.~where the first time derivative of the 
argument of the phase of the integrand vanishes. For generic 
precessing systems, however, the argument of the phase of the generalized 
Fourier integrand will encounter catastrophes~\cite{Berry1980257}, 
where multiple stationary points coalesce, and the second 
time derivative of the phase of the integrand vanishes together with the first 
derivative. Such catastrophes 
introduce mathematical singularities in the SPA Fourier amplitude 
that render the SPA ill suited for the construction of frequency-domain 
templates for precessing systems. 

One way around these precession-induced catastrophes is through the method of 
uniform
asymptotic expansion~\cite{Berry1980257,Bender}. When the phase oscillates, as 
it does here,
the Fourier integrand can be expanded in a Bessel 
series~\cite{PhysRevE.64.026215,Klein:2013}. The idea
is to rewrite the phase of the integrand by expanding the oscillatory terms that 
depend on the spin and angular momentum, and thus on 
trigonometric functions of the precession phases, in an infinite Bessel 
series. Although 
the full series is still problematic from a catastrophe theory standpoint, each 
term in the series has a well-behaved SPA.
One then avoids catastrophes by truncating  the infinite
sum at a finite order, which is justified because the contribution from the 
Bessel functions that 
multiply the amplitude at high orders in the sum are very small.
The resulting frequency-domain waveform remains highly accurate 
and faithful~\cite{Klein:2013}, but the need to introduce Bessel series makes 
them relatively computationally expensive, which partially defeats the purpose 
of constructing them in the first place. 

In this paper, we propose a new approach, the method of shifted 
uniform asymptotics (SUA) that resums the Bessel 
function
series generated by the uniform asymptotic expansion, resulting in a sum of 
time-shifted precession modulations. The SUA allows us to construct 
frequency-domain waveforms
for generic  precessing systems without the need to evaluate Bessel expansions. 
This 
method takes as input any closed-form or numerical time-domain solution to the 
orbital angular momentum, the orbital phase, and the orbital frequency, and 
outputs a remarkably simple, frequency-domain representation of the GW response. 
In essence, 
this approach \emph{resums} the Bessel expansions described 
in Ref.~\cite{Klein:2013}, using the generic Taylor expansion of a shifted 
function; the result is a frequency-domain waveform that resembles the usual 
SPA  result,  but with a correction that consists of a sum of amplitudes shifted 
with 
respect to the SPA stationary time. 
The SUA frequency-domain template is simple to compute, 
computationally inexpensive and highly 
accurate relative to time-domain 
waveforms obtained through a DFT.

As an example, we apply the SUA method to a numerical, 
time-domain 
solution for generic precessing binaries with arbitrary spin magnitude and 
orientation. In particular, we adopt the \texttt{SpinTaylorT4}\footnote{Here we 
are using the naming conventions of the LIGO Algorithm Library, {\tt 
https://www.lsc-group.phys.uwm.edu/ daswg/projects/lalsuite.html}.} model as 
the 
time-domain solution, i.e.~the numerical solution to the 2PN expanded 
precession 
and 3.5PN orbital equations\footnote{An expression is said to be $N$PN order 
accurate 
when it contains
relative uncontrolled remainders of ${\cal{O}}(v^{2N+1})$.}. This model yields 
a 
time series solution for the orbital 
angular momentum and phase, as well as a time series for the waveform itself. 
Using the former, we compute the SUA, frequency-domain 
representation to the \texttt{SpinTaylorT4} waveform and compare it to the 
DFT of the \texttt{SpinTaylorT4} time series. In the rest of this paper, we 
will refer to the frequency-domain SUA version of the \texttt{SpinTaylorT4} 
waveforms as \texttt{SpinTaylorT4Fourier} to distinguish them from the
usual time-domain \texttt{SpinTaylorT4} waveforms. 
Recall that we use the word ``waveform'' to refer to frequency series as is 
customary in the GW data analysis community.
We find excellent agreement between these two waveforms, 
with unfaithfulness\footnote{The unfaithfulness is a particular noise-weighted 
cross-correlation between two waveforms in the frequency domain, 
minimized \emph{only} over unphysical parameters. 
An unfaithfulness of zero implies perfect agreement between the two waveforms.} 
of the order of $10^{-5}$. The SUA waveforms, however, are found 
to be much faster to generate than 
time-domain templates. All of the above is implemented 
and 
carried out 
in the \texttt{lalsimulation} open-source package of the LIGO Scientific 
Collaboration, including the SUA version of the 
\texttt{SpinTaylorT4} and \texttt{SpinTaylorT2} templates, which we called 
\texttt{SpinTaylorT4Fourier} and \texttt{SpinTaylorT2Fourier} respectively. 
A definition of those PN flavors can be found in 
e.g.~\cite{buonanno2009,Nitz2013}.

\begin{figure}[ht]
 \includegraphics[width=0.95\columnwidth]{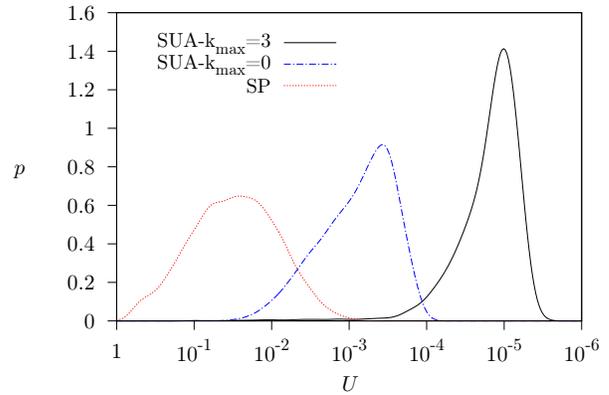}
 \caption{\label{BHBH3}Distributions of the unfaithfulness between time-domain 
\texttt{SpinTaylorT4} waveforms and (i) the corresponding frequency-domain SUA 
waveforms, truncating the sum of shifted amplitudes in 
Eq.~\eqref{finalWF} at the zeroth term 
(dot-dashed blue)
or at the third term (solid black); 
and (ii) simple precession single-spin waveforms (dotted 
red)~\cite{Hannam:2013}. The systems are 
highly spinning 
black hole binaries with random masses in $(5,20) M_\odot$, 
random dimensionless spin magnitudes in $(0,1)$, and 
random orientations. Observe that the SUA waveforms show 
excellent agreement 
(extremely low unfaithfulness) with time-domain 
\texttt{SpinTaylorT4} 
waveforms.}
\end{figure}

Figure~\ref{BHBH3} shows how faithful the SUA waveforms are. 
This figure shows mismatch distributions between 
time-domain \texttt{SpinTaylorT4} 
waveforms and either 
\begin{itemize}
\item[(i)] The  \texttt{SpinTaylorT4Fourier} SUA waveforms constructed with \texttt{SpinTaylorT4} 
time-domain 
solutions to the precession equations and with the sum of shifted amplitudes 
in Eq.~\eqref{finalWF}
truncated at either the zeroth term (dot-dashed blue) or the third 
term (solid black); 
\item[(ii)] the single-spin, simple precession approximation of 
Ref.~\cite{Hannam:2013} (dotted red), constructed by solving the precession
equations approximately, assuming one of the two compact objects is not 
spinning.
\end{itemize}
The time-domain \texttt{SpinTaylorT4} waveform and the frequency-domain
\texttt{SpinTaylorT4Fourier} waveform include subdominant PN harmonic
corrections. The single-spin waveform contains only the $(\ell,|m|) = 
(2,2)$ amplitude. The distributions are created through 
Monte Carlo sampling 
over system parameters for a binary black hole inspiral, with mass range in 
$(5,20) M_{\odot}$, spin angular momentum magnitude in $(0, m_{A}^{2})$, 
where $m_{A}$ is the individual BH mass, and random spin orientations. Observe 
that the single-spin templates have relatively 
poor agreement with time-domain \texttt{SpinTaylorT4} waveforms, while the 
SUA templates show excellent agreement. Observe also that 
retaining only the zeroth order term in the sum of shifted amplitudes is not 
sufficient 
to ensure $> 99\%$ faithfulness.

The remainder of this paper presents the details of the SUA 
method and of the comparison described above. 
Section~\ref{derivation} derives the SUA method. 
Section~\ref{comparison} establishes the validity of the method by applying it 
and comparing it to time-domain \texttt{SpinTaylorT4} templates. 
Section~\ref{conclusion} concludes and points to future work. 

Throughout this paper, we use geometric units where $G=c=1$, as well as the 
following conventions and notation:
\begin{itemize}
\item Three-dimensional vectors are written in boldface and unit vectors carry 
a 
hat over them, e.g.~$\bm{A} = (A_x, A_y, A_z)$, with norm $A = |\bm{A}|$, and 
unit vector $\uvec{A} = \bm{A}/A$.
\item Total time derivatives are denoted with a dot, e.g. $\dot{f} = df/dt$. 
\item The individual masses of the binary system components are $m_1$ and 
$m_2$, with $m_1 
> m_2$, the total mass is $M = m_1 + m_2$, the dimensionless  individual masses 
are $\mu_A = m_A/M$, $A \in \{1,2\}$, and the symmetric mass ratio is $\nu = 
\mu_1 \mu_2$.
\item $\bm{L}$ is the Newtonian orbital angular momentum.
\item $\bm{S}_A$, $A \in \{1,2\}$ are the individual spin angular momentum 
vectors, $\bm{a}_A = \bm{S}_A/(m_A M)$ are dimensionless spin 
vectors, and $\chi_A = S_A / m_A^2$ are the individual spin parameters.
\item $\uvec{N}$ is the sky localization vector from the detector to the source.
\end{itemize}

\section{The Shifted Uniform Asymptotics Method}
\label{derivation}

In this section, we derive the SUA method to obtain the 
frequency-domain representation of a time-domain waveform for generic 
precessing 
inspirals that avoids catastrophes.  For concreteness, we focus on GWs emitted 
during the quasicircular inspiral of compact objects with generic spin 
magnitudes and orientations, such that the orbital plane is undergoing generic 
precession. We study the Fourier transform of the response of advanced, 
ground-based detectors to such GWs, assuming an L-shaped detector in the 
long wavelength approximation. Our results, of course, could easily be extended 
to other 
detector configurations, such as eLISA~\cite{elisa}.


\subsection{Preliminaries}

The time-domain response function can be given 
as~\cite{Apostolatos:1994mx,blanchet-review,abiq}
\begin{align}
h(t) &= F_+(t) h_+(t) + F_\times(t) h_\times(t)\,,
\end{align}
where the time-domain GW plus- and cross-polarizations can be written as
\begin{align}
h_{+,\times} &= \sum_{n \geq 0}  \mathcal{A}_{+,\times}^{(n)}(\iota) \; e^{- i 
n(\phi_C + \phi_T)} + \mbox{c.c.}\, , \label{hoft}
\end{align}
with ${\rm{c.c.}}$~the complex conjugate, $n \in \mathbb{N}$ the harmonic 
number, $\iota \equiv \arccos(\uvec{L} \cdot \uvec{N})$ the inclination angle, 
$\phi_C$ the carrier GW phase, $\phi_T$ the \emph{Thomas phase}, which 
accounts for the precession of the greater axis of the projection of the 
orbital 
plane onto the sky inside the orbital plane as $\bm{L}$ precesses. The antenna 
or beam pattern functions are
\begin{align}
 F_+(\theta_N, \phi_N, \psi_N) &= \frac{1}{2} \left( 1 + \cos^2 \theta_N
\right) \cos 2\phi_N \cos
2\psi_N \nonumber\\
&- \cos \theta_N \sin 2\phi_N \sin 2\psi_N, \label{eq:Fplus}\\
 F_\times(\theta_N, \phi_N, \psi_N) &= F_+(\theta_N,\phi_N,\psi_N-\pi/4),
\label{eq:Fcross}
\end{align}
where $(\theta_N,\phi_N)$ are spherical angles that label the position of the 
binary in a frame tied to the detector, with $\uvec{x}$ and $\uvec{y}$ unit 
vectors along the arms of the detector, $\uvec{z} = \uvec{x} \times  \uvec{y}$, 
and $\psi_N$ the polarization angle defined through
\begin{equation}
 \tan \psi_N = \frac{\uvec{L}\cdot\uvec{z} - (\uvec{L}\cdot\uvec{N})
(\uvec{z}\cdot\uvec{N})}{\uvec{N}\cdot( \uvec{L}\times\uvec{z} )}\,.
\label{eq:psiN}
\end{equation} 


\subsection{Separation of scales}
A key element of the SUA method is the use of multiple-scale 
analysis~\cite{Bender}. In order to illustrate this, consider the equations of 
motion during the inspiral in the so-called \texttt{SpinTaylorT4} 
form~\cite{blanchet-review,racinebuonannokidder,Favata:2009,blanchet3PN,
Bohe:2013cla,PhysRevD.12.329, Bohe:2012mr}:
\begin{align}
 M \dot{\phi}\sub{orb} &= \xi^3 , \label{orbtime}\\
 \phi_C &= \phi\sub{orb} - (6 - 3 \nu \xi^2) \xi^3 \log\xi, \label{phic} \\
 M \dot{\xi} &=  \xi^9 \sum_{n = 0}^{N} b_n \xi^n, 
\label{rrtime} \\
\dot{\phi}_T &= \frac{\cos\iota}{1 - \cos^2\iota} \left( \uvec{L} \times 
\uvec{N} \right) \cdot
 \duvec{L}, \\
 M\duvec{L} &= - \xi^6 \left( \bm{\Omega}_1 + 
\bm{\Omega}_2 \right), \label{prectimeL} \\
 M\dvec{a}_1 &= \mu_2 \xi^5 \bm{\Omega}_1, \label{prectimes1} \\
 M\dvec{a}_2 &= \mu_1 \xi^5 \bm{\Omega}_2, \label{prectimes2}
\end{align}
where $\xi = \mathcal{O}(v/c) \ll 1$ is a PN expansion parameter and 
$\phi\sub{orb}$ is the orbital phase. The coefficients $b_n = 
\mathcal{O}(\xi^0)$ and $\bm{\Omega}_i = \mathcal{O}(\xi^0)$ are given in 
Appendix~\ref{app:eqmot}. The vectors $\bm{\Omega}_i$ satisfy $\bm{\Omega}_i 
\perp \bm{a}_i$ 
and 
$(\bm{\Omega}_1 + \bm{\Omega}_2 ) \perp \uvec{L}$. From these equations, one 
can 
 see that three separate time scales arise naturally: the orbital 
timescale $T\sub{orb} = \mathcal{O}(\xi^{-3})$ defined through 
Eq.~\eqref{orbtime}; the radiation-reaction time scale $T\sub{r.r.} = 
\mathcal{O}(\xi^{-8})$, defined through Eq.~\eqref{rrtime}; and the precession 
time scale $T\sub{prec} = \mathcal{O}(\xi^{-5})$, defined through 
Eqs.~(\ref{prectimeL})-(\ref{prectimes2}). We thus have a natural separation of 
scales, $T\sub{r.r.} \gg T\sub{prec} \gg T\sub{orb}$, that is tailor made for 
a multiple-scale analysis treatment. 

With this separation of scales in mind, we can rewrite the time-domain response 
as
\begin{align}
\label{eq:h(t)-pre}
 h(t) = \sum_{n,k,m} \mathcal{A}_{n,k,m} e^{-i(n \phi_T + k \iota + m \psi_N)} 
e^{- i n \phi_C}  + \mbox{c.c.}\,,
\end{align}
where we note that $\phi_C$ varies on the orbital time scale while $\phi_T$, 
$\iota$, and $\psi_N$ vary on the precession time scale. Furthermore, we can 
note 
from 
Eq.~\eqref{prectimeL} that it is natural to expand $(\phi_{T},\iota,\psi_{N})$ 
in a Fourier series on the precession time scale, i.e. with amplitudes $A_{j}$ 
and phases $\beta_{j}$ that satisfy $\dot{\beta}_j = 
\mathcal{O} (\xi^5)$ and $A_j = \mathcal{O}(\xi)$. We can then rewrite 
Eq.~\eqref{eq:h(t)-pre} as 
\begin{align}
h(t) = \sum_{n,\bm{A}} \mathcal{A}_{n,\bm{A}} e^{-i \sum_j A_j \sin 
\beta_j} 
e^{- i n \phi_C}  + \mbox{c.c.}\, ,
\end{align}
where $\bm{A}$ is an abstract amplitude vector that represents the dependence 
of 
each harmonic on the precession phases $\beta_j$ and varies on the 
radiation-reaction time scale.

Separation of scales has revealed that the calculation of the Fourier transform 
of $h(t)$ is equivalent to transforming another function $H(t)$ defined via
\be
 H(t) = H_0(t) \;  e^{- i \delta\phi(t)}\,,
\ee
with ``background'' response
\be
H_0(t) = \mathcal{A} \; e^{-i n \phi_C}\,,
\ee
and a precession correction
\be
\delta\phi = \sum_j A_j \sin \beta_j\,,
\ee
that satisfy
\begin{align}
\dot{\phi}_C &\sim \mathcal{O} (\xi^3), \qquad
\dot{\beta}_j \sim \mathcal{O} (\xi^5), \\
\mathcal{A} &\sim \mathcal{O} (\xi^2), \qquad
A_j \sim \mathcal{O} (\xi),
\end{align}
where $\mathcal{A}$, $A_j$, $\dot{\phi}_C$, and $\dot{\beta}_j$ all vary on the 
radiation reaction time scale. We will derive the SUA method 
using this toy problem first, and
then generalize it to the actual GW response. 


\subsection{Bessel expansion}
We begin by expanding the precession correction in a Bessel 
series~\cite{Klein:2013}:
\begin{align}
 H(t) &= \sum_{\bm{k} \in \mathbb{Z}^\mathbb{N}} \mathcal{A} \left[ \prod_m
J_{k_m}(A_m) \right] e^{-i(n \phi_C + \sum_j k_j \beta_j)}\,.
\end{align}
We can now Fourier transform $H(t)$ using the stationary phase approximation 
because the second derivative of 
the argument of the imaginary exponential is always negative, and thus, it 
avoids catastrophes, i.e.~times at which this second derivative would vanish. 
Note though that, in principle, the catastrophes are actually still there when 
one
includes arbitrarily large values of $k_{j}$, since then one could have 
$\sum_j k_j \ddot{\beta}_j \sim n \ddot{\phi}_C$. For such large values of the 
sum index,
however, the corresponding Bessel amplitudes suppress the divergences. Using 
the 
SPA, one finds
\begin{align}
 \tilde{H}(f) &\approx \sum_{\bm{k} \in \mathbb{Z}^\mathbb{N}} 
\mathcal{A}(\tk) \left[ \prod_m 
J_{k_m}[A_m(\tk)] \right] \nonumber\\
&\times \sqrt{\frac{2\pi}{n 
\ddot{\phi}_C(\tk) + \sum_j k_j \ddot{\beta}_j(\tk)}} \nonumber\\
&\times e^{i[2\pi f \tk - n \phi_C(\tk) - \sum_j k_j 
\beta_j(\tk) - \pi/4]}\,, \label{Htilde}
\end{align}
where the stationary points $\tk$ are defined via the condition
\be
2 \pi f = n \dot{\phi}_C (\tk) + \sum_j k_j \dot{\beta}_j (\tk). 
\label{foftk}
\ee

Let us now reexpress the full Fourier transform in a 
\emph{product decomposition} of the form $\tilde{H}(f) = \tilde{H}_0(f) \; 
\tilde{H}\sub{corr}(f) $, 
where $\tilde{H}_0(f)$ is the Fourier transform of the background response
\begin{align}
 \tilde{H}_0(f) &\approx \mathcal{A}(t_0) \sqrt{\frac{2\pi}{n 
\ddot{\phi}_C(t_0)}} e^{i[2\pi f 
t_0 - n \phi_C 
(t_0) - \pi/4]}\,,
\label{foft0}
\end{align}
again computed in the SPA because $\ddot{\phi}_C > 0$, where $t_{0}$ is the 
stationary point defined through the condition $2 \pi f = n \dot{\phi}_C(t_0)$.
Note that $\tilde{H}\sub{corr}(f)$ is not necessarily the Fourier 
transform of a known time-domain signal, and carries a tilde only to stress 
that it is a Fourier-domain quantity.

To do so, we keep only leading PN order terms in the amplitude, and neglect any 
factors of $\mathcal{O}(\xi)$ or higher in the phase. 
First, we Taylor expand the stationary phase conditions to obtain
\begin{align}
 \tk &= t_0 + \Delta \tk, \label{tk}\\
 \Delta \tk &=  - \frac{1}{n\ddot{\phi}_C(t_0)} \sum_j k_j \dot{\beta}_j(t_0) + 
\mathcal{O} (\xi^{-4})\,.
\end{align}
Expanding the amplitude of Eq.~\eqref{Htilde} to leading PN order, we then find
\begin{align}
\tilde{H}(f) &\approx \mathcal{A}(t_0) \sqrt{\frac{2\pi}{n\ddot{\phi}_C(t_0)}} 
\sum_{\bm{k} \in  \mathbb{Z}^\mathbb{N}} \left[ \prod_m  J_{k_m}[A_m(t_0)] 
\right]  e^{ i \Psi_{\bm{k}}},
\end{align}
where the Fourier phases $\Psi_{\bm{k}}$ can be expanded using Eq.~\eqref{tk} 
to 
find
\begin{align}
 \Psi_{\bm{k}} &= 2\pi f t_0 - n \phi_C(t_0) -  \frac{\pi}{4} - \sum_j k_j 
\beta_j(t_0) 
 \nonumber\\
&+ 
\frac{1}{2 n\ddot{\phi}_C(t_0)} \left[\sum_j k_j \dot{\beta}_j(t_0)\right]^2 + 
\mathcal{O}(\xi).
\end{align}
Combining these results, the precession correction to the Fourier transform is 
then 
simply
\begin{align}
\tilde{H}\sub{corr}(f) &= \sum_{\bm{k} \in 
\mathbb{Z}^\mathbb{N}} \left[ \prod_m 
J_{k_m}[A_m(t_0)] \right] e^{i \Delta\Psi_{\bm{k}}},
\end{align}
with the precession phase correction
\be
\Delta\Psi_{\bm{k}} = - \sum_j k_j \beta_j(t_0) + 
\frac{1}{2} T^2 \left[\sum_j k_j \dot{\beta}_j(t_0) \right]^2, 
\label{eq:DeltaPsik}
\ee
with the new time scale $T = [n \; \ddot{\phi}_C(t_0) ]^{-1/2}$.

\subsection{Bessel resummation}

The product decomposed result obtained above is similar to that 
of~\cite{Klein:2013}, but inefficient in practice because of the large number 
of 
terms that must be kept in the Bessel expansion.  
However, 
all of these Bessel terms can be resummed. To do so, let us first note 
that 
\begin{align}
 \partial_t \left(A_j \sin \beta_j\right) &= \dot{\beta}_j A_j \left[ \cos 
\beta_j + \mathcal{O}(\xi^3) \right],
\end{align}
which then implies
\begin{align}
 \partial_t^q \left[f( \delta\phi)\right] &= \left[ 1 + \mathcal{O} (\xi^3) 
\right]\left[\sum_j \dot{\beta}_j \partial_{\beta_j} \right]^q  
f(\delta\phi)  .
\label{eq:partialt-relation}
\end{align}
This relation allows us to simplify the precession correction of the Fourier 
transform to
\begin{widetext}
\begin{align}
\tilde{H}\sub{corr}(f) &= \sum_{p \geq 0} \frac{\left(i T^2\right)^p}{2^p p!} 
\sum_{\bm{k} \in \mathbb{Z}^\mathbb{N}} \left(\sum_\ell k_\ell \dot{\beta}_\ell 
\right)^{2p} 
\left[  \prod_m J_{k_m}(A_m) \right] e^{-i \sum_j k_j  \beta_j}  
\\
&= \sum_{p \geq 0} \frac{\left(i T^2\right)^p}{2^p p!} 
\left(i\sum_\ell \dot{\beta}_\ell \partial_{\beta_\ell} \right)^{2p} 
\sum_{\bm{k} \in  \mathbb{Z}^\mathbb{N}} 
\left[ \prod_m  J_{k_m}(A_m) \right] e^{-i \sum_j k_j \beta_j} 
\\
&= \sum_{p \geq 0} \frac{\left(-i T^2\right)^p}{2^p p!} 
 \left(\sum_\ell \dot{\beta}_\ell \partial_{\beta_\ell} \right)^{2p} e^{- i 
\delta\phi} 
= \sum_{p \geq 0} \frac{\left(-i T^2\right)^p}{2^p p!}  
\partial_t^{2p} e^{- i \delta\phi}  \left[ 1 + \mathcal{O} (\xi^3) \right]. 
\label{Hcorr_intermediate}
\end{align}
\end{widetext}
where in the first equality we have Taylor expanded the exponential 
produced by the second term in Eq.~\eqref{eq:DeltaPsik}, in the second equality 
the $k_{\ell}$ is replaced by the $\partial_{\beta_{\ell}}$ derivative 
acting on $\exp(-i \sum_{j} k_{j} \beta_{j})$, in the third equality we 
resummed 
the Bessel expansion of $\exp(-i \delta \phi)$ and in the last equality we used 
Eq.~\eqref{eq:partialt-relation}.

With this at hand, we can now use the shift relation $F(x+h)=e^{h \frac{d}{dx}}F(x)$ in the form
\begin{align}
 f(t_0 + kT) &= \sum_{p \geq 0} \frac{(k T)^p}{p!} \partial_t^p f(t_0)\,,
\end{align}
which is the definition of a Taylor expansion. This expression allows us to 
write the following 
double shift relation
\begin{align}
e^{-i \delta\phi(t_0 + kT)} + e^{-i \delta\phi(t_0 - kT)} = 2\sum_{p \geq 0} 
\frac{(k 
T)^{2p}}{(2p)!} \partial_t^{2p} e^{-i \delta\phi(t_0)}\,, \label{doubleshift}
\end{align}
which we can use to rewrite the precession correction of the Fourier transform 
as
\begin{align}
 \tilde{H}\sub{corr}(f) &\approx \sum_{p = 0}^{k\sub{max}} \frac{\left(-i 
T^2\right)^p}{2^p p!}  
\partial_t^{2p} e^{- i \delta\phi(t_0)} 
\label{Hcorr1}
\\
&= \sum_{k = 0}^{k\sub{max}} \sum_{p = 0}^{k\sub{max}} a_{k,k\sub{max}} 
\frac{(k 
T)^{2p}}{(2p)!} \partial_t^{2p} 
e^{-i \delta\phi(t_0)} \\
&\approx \frac{1}{2} \sum_{k = 0}^{k\sub{max}} a_{k,k\sub{max}} \left[ e^{-i 
\delta\phi(t_0 + kT)} + e^{-i \delta\phi(t_0 - kT)} \right]\,,
\label{Hcorr2}
\end{align}
where the $a_{k,k\sub{max}}$ are constant coefficients.
The first equality is simply Eq.~\eqref{Hcorr_intermediate} truncated 
at some integer order $k\sub{max}$. The second 
equality is true provided that the coefficients $a_{k,k\sub{max}}$ satisfy
\begin{align}
 \frac{(-i)^p}{2^p p!} = \sum_{k = 0}^{k\sub{max}} a_{k,k\sub{max}} 
\frac{k^{2p}}{(2p)!}, 
\label{ak_def}
\end{align}
with $p \in \{0, \ldots, k\sub{max}\}$. This is an easily solvable linear 
system of $k\sub{max}+1$ equations for $k\sub{max}+1$ variables.
The third equality Eq.~\eqref{Hcorr2}, is established through 
Eq.~\eqref{doubleshift} truncated at order $k\sub{max}$.

Another way of deriving Eqs.~\eqref{Hcorr2} and~\eqref{ak_def} from 
Eq.~\eqref{Hcorr_intermediate} can shed some light on the meaning of the 
constants $a_{k,k\sub{max}}$ and is presented in appendix~\ref{app:stencils}.

\subsection{The SUA Fourier response}

We can now go back to $h(t)$ and compute its Fourier transform 
$\tilde{h}(f)$. We can rewrite Eq.~\eqref{hoft} as
\be
 h(t) = \sum_n \mathcal{A}_n(t) e^{-i n \phi_C} + \mbox{c.c.}\, , \label{hoft2} 
\ee
with the precession-dependent amplitude
\be
\mathcal{A}_n(t) = e^{- i n \phi_T(t)} \left\{ F_+(t) \mathcal{A}_+^{(n)} 
[\iota(t) ] + F_\times(t) \mathcal{A}_\times^{(n)} [\iota(t) ] \right\}.
\ee
By analogy with the toy problem, the Fourier transform is then approximately
\begin{align}
\tilde{h}(f) &\approx \sum_n \sqrt{\frac{2 \pi}{n 
\ddot{\phi}\sub{orb}(t_{0,n})}} 
\Bigg\{ \sum_{k=0}^{k\sub{max}} \frac{a_{k,k\sub{max}}}{2} [ 
\mathcal{A}_n(t_{0,n} + k T_n) 
\nonumber\\
&+  
\mathcal{A}_n(t_{0,n} - k T_n) ] \Bigg\} e^{i[ 2 \pi f t_{0,n} - n 
\phi_C(t_{0,n}) - \pi/4]}\,,
\label{finalWF} 
\end{align}
where the stationary points are simply given by $2 \pi f = n 
\dot{\phi}\sub{orb}(t_{0,n})$, and $T_n = [n 
\ddot{\phi}\sub{orb}(t_{0,n})]^{-1/2}$, with constants $a_{k,k\sub{max}}$ that 
satisfy the 
linear system of equations defined by Eq.~\eqref{ak_def} with $p \in \{0,1, 
\ldots , k\sub{max}\}$. In the above expression, we replaced time derivatives 
of 
$\phi_C$ by time derivatives of $\phi\sub{orb}$, because $\dot{\phi}_C 
- \dot{\phi}\sub{orb} = \mathcal{O}(\xi^{11}) \ll \dot{\phi}\sub{orb} 
=\mathcal{O}(\xi^3)$, as established from Eq.~\eqref{phic}.

The Fourier transform of any time-domain GW response function is given by 
Eq.~\eqref{finalWF}, which only requires knowledge of $\xi(t)$, 
$\phi\sub{orb}(t)$, and $\uvec{L}(t)$. Notice that one does not need to specify 
whether the latter are derived analytically or numerically, or the waveform in 
a 
particular frame, or a particular approximant when solving the equations of 
motion. The SUA Fourier response only requires that time scales 
separate, a valid assumption that breaks down only close to plunge.

Interestingly, 
if we retain only the lowest order solution in Eq.~\eqref{finalWF} with 
$k\sub{max} = 
0$, then $a_0 = 1$ and one recovers a waveform similar to the one proposed by 
Lang and 
Hughes in~\cite{Lang:1900bz}. Their waveform is in fact the $k\sub{max}=0$ SUA 
waveform, with (i) computing $\uvec{L}(t)$ by numerically integrating
Eqs.~(\ref{prectimeL})-(\ref{prectimes2}), (ii) computing the relations $t(f)$ 
and 
$\phi\sub{orb}(f)$ by analytically integrating Eq.~\eqref{orbtime} 
through a PN expansion,
(iii) restricting the amplitudes $\mathcal{A}_\pm^{(n)}$ to leading order, 
and (iv) replacing $\phi_C$ by $\phi\sub{orb}$ which is justified when one uses 
leading-order amplitudes.
The calculation described above provides a 
mathematical 
justification for this prescription and it includes the necessary corrections 
to extend it. However, a crucial difference between it and the $k\sub{max}=0$ 
solution that we used in the following section is that we used a numerical 
solution for the relations $t(f)$ and 
$\phi\sub{orb}(f)$ instead of an analytical one, which {\em significantly} improves the
accuracy of the waveforms.

\section{Implementation and Validation}
\label{comparison}

In this section, we will use the SUA method to construct a 
particular
example that we can then validate through certain data analysis measures. 
For this particular example, we will use the \texttt{SpinTaylorT4} model to 
construct the 
time-domain evolution of the orbital angular momentum and phase, which will
serve as input into the SUA method. We will then validate these 
templates
against time-domain \texttt{SpinTaylorT4} waveforms. The comparisons will be 
done
through the faithfulness, one minus the unfaithfulness measure shown in the 
Introduction. 
We stress however that the SUA method is \emph{generic};
one can apply this method to any analytic or numeric evolution of the orbital 
angular momentum
and phase. 

\subsection{Preliminaries}

The data analysis measure we will use in this section to compare templates will 
be the 
faithfulness $F$ between waveforms $h_1$ and $h_2$, defined as
\begin{align}
\rm{F}(h_1, h_2) &= \max_{\lambda^{a}_{\up}} \int_{f\sub{min}}^{f\sub{max}} 
\frac{\hat{h}_1(f) 
\hat{h}_2^*(f)}{S_n(f)} df,
\end{align}
where $S_n(f)$ is the noise spectral density of the detector we are 
considering, $f\sub{min}$ and $f\sub{max}$ are detector-dependent frequency 
cutoffs, and the normalized waveform $\hat{h}_{A}(f)$ is
\begin{align}
 \hat{h}_{A}(f) &= \left[\int_{f\sub{min}}^{f\sub{max}} \frac{| 
\tilde{h}_{A}(f)|^2}{S_n(f)} df 
\right]^{-1/2} \tilde{h}_A(f),
\end{align}
for $A = 1$ or $2$ and where $\tilde{h}_{A}(f)$ is the Fourier transform of 
$h_{A}(t)$. For concreteness, we here focus on aLIGO, with $S_{n}$ given by an 
analytical fit found in~\cite{cornishsampson} to the zero-detuned, 
high-power spectral noise density projected for aLIGO~\cite{aligonoise} and 
$(f\sub{min},f\sub{max}) = (10,10^{4})$~Hz.

The faithfulness is maximized over all unphysical parameters 
$\lambda^{a}_{\up}$. 
For SPA frequency-domain waveforms, these parameters correspond to a global 
time 
shift and 
a global orbital phase shift, which appear as constants of integration when 
computing the Fourier phase. Maximization over these parameters is 
necessary when comparing models for which the frequency and phase evolutions 
$\omega\sub{orb}(t)$ and $\phi\sub{orb}(t)$ differ.
For the SUA templates and the seed time-domain 
waveform, 
these two parameters correspond to the initial conditions of the evolution 
equations; therefore, 
by construction, the same parameters lead to the same evolution in 
both and one does not need to 
maximize 
the faithfulness
with respect to them. 

We chose to use the faithfulness instead of the fitting factor, defined 
similarly but maximized over \emph{all} 
parameters, because our waveforms are designed for fast parameter estimation. 
The fitting factor is a better-suited measure for waveforms aimed at detection, 
since in that case one does not worry about parameter biases. 
Notice, of course, that the faithfulness between two identical templates is 
simply unity, 
i.e.~$\rm{F}(h_{1}, h_{1}) = 1 = \rm{F}(h_{2}, h_{2})$, 
while the unfaithfulness
used in the Introduction is simply defined as $\rm{U} = 1 - \rm{F}$. 

An often cited bound for the fitting factor or similar measures like 
the faithfulness is 0.97. The origin of it is 
that the average fitting factor for a waveform needs to be 
$0.9^{1/3} \approx 0.965$ for it to recover $90\%$ of the total number of 
signals in an experiment~\cite{apostolatos1995,owensathya}. However, the 
requirements on the faithfulness for parameter estimation (PE) studies are 
different, and depend on the signal-to-noise ratio (SNR). A sensible requirement 
for a PE study is to ask that the systematic (or mismodeling) error coming from 
using an approximate waveform is lower than the statistical error coming from 
the noise in the data. While the former is SNR independent, the latter does 
depend on the SNR, and therefore the faithfulness requirement on a waveform for 
a PE study is SNR dependent: the faithfulness requirement scales like the 
inverse SNR squared $1-F\sub{req} \sim 1/\text{SNR}^2$.

The calculation of the faithfulness requires the choice of at least two 
templates. One of them will always be the time-domain \texttt{SpinTaylorT4} 
waveform. The other will be one of the following:
\begin{itemize}
\item[(i)] \emph{the} \texttt{SpinTaylorT4Fourier} \emph{SUA, frequency-domain templates}  
computed with the \texttt{SpinTaylorT4} numerical evolution for the orbital 
angular momentum 
and phase for values of $k\sub{max}$ equal to 
every integer between $0$ and $10$;
\item[(ii)] \emph{the small-spins, double precessing waveforms} 
(DP)~\cite{Chatziioannou:2013}, computed by solving the precession equations 
analytically assuming the individual dimensionless spin magnitudes are 
much smaller than unity, i.e.~$\chi_{A} \ll 1$;
\item[(iii)] \emph{the single-spin, simple precession waveforms} (SP), also 
known as the \texttt{PhenomP} model~\cite{Hannam:2013}, constructed 
by mapping generic double-spin systems onto single-spin ones.
\end{itemize}

The PN order to which each of these templates is valid is a tricky issue. 
The time-domain \texttt{SpinTaylorT4} waveform, in principle, contains all 
valid terms only up to 2PN order; point-mass terms and spin-orbit terms are 
known
to higher PN order (see 
e.g.~\cite{Bohe:2012mr,Marsat:2013wwa,Jaranowski:2012eb,Jaranowski:2013lca,
Damour:2014jta}), but not all of them are used in the \texttt{SpinTaylorT4} 
model.
In the DP and SP 
templates, we choose
to limit the accuracy of the PN Fourier phase to 3.5PN order. Although in 
principle some higher PN
order terms could be kept, they would not be consistent with full general 
relativity beyond 3.5PN. 

The waveform templates can differ also in the number of terms kept in the 
wave amplitude. The restricted PN approximation, or restricted waveforms 
(RWF) for short, consists of keeping only the leading, PN order term in the 
wave amplitude, i.e.~the $n=2$ harmonic in a multipolar decomposition for 
a quasicircular inspiral, without PN corrections to it. Full waveforms (FWF) 
consist of waveforms constructed with keeping as many PN corrections in the 
wave amplitude as 
possible. 
In what follows, the time-domain \texttt{SpinTaylorT4} waveforms will include 
these higher PN order harmonics, while the SP model will keep only the 
$(\ell,|m|) = (2,2)$ term in the amplitude. The SUA and DP models will be 
studied both with 
RWFs and FWFs.

When comparing frequency-domain waveforms to time-domain 
ones that end abruptly, one needs to pay attention to spectral leakage. 
Time-domain waveforms ending abruptly are effectively multiplied by a 
Heaviside function, and thus their Fourier transform will be convoluted 
with the Fourier transform of it. This results in unwanted oscillations near 
the beginning and the end of the signal. To reduce that effect, we use a Tukey 
window on the signal, defined by

\begin{align}
 W(t) &= \left\{ 
 \begin{array}{l l}
 0, & t < t_1 \\
 \sin^2 \left( \frac{t - t_1}{t_2 - t_1} \right), & t_1 \leq t < t_2 \\
  1, & t_2 \leq t < t_3 \\
 \sin^2 \left( \frac{t_4 - t}{t_4 - t_3} \right), & t_3 \leq t < t_4 \\
 0, & t_4 \leq t
  \end{array}
\right. .
\end{align}

This will reduce leakage but will also reduce the available power.
A good compromise for $t_1, \ldots, t_4$ is for $\sim10$~cycles to occur 
between $t_1$ and $t_2$, and as many between $t_3$ and $t_4$. If that condition 
is satisfied, then the window is a slowly varying function from the point of 
view of the SPA, and can therefore easily be taken into account in the 
frequency-domain waveforms.
We chose $t_1$ and $t_2$ to be too small to have an influence on the 
frequency-domain waveforms, i.e.~$n\sub{max} \dot{\phi}\sub{orb}(t_1, t_2) < 
10$~Hz, where $n\sub{max}$ is the highest harmonic number. We chose $t_3$ and 
$t_4$  so that $\xi(t_3) = 15^{-1/2}$ and $\xi(t_4) = 6^{-1/2}$. The 
frequency-domain waveforms also need a $t(f)$ relation in order to use the 
window. We chose to use the post-Newtonian relations $t_n(f)$ to an order 
consistent with the phase where $t_n(f)$ is the time at which harmonic $n$ 
emits radiation at frequency $f$, so that each harmonic is multiplied by 
$W[t_n(f)]$.

The calculation of the faithfulness also requires a choice of systems to study. 
We here
focus on the following:
\begin{enumerate}
 \item[(a)] Highly spinning neutron star--neutron star systems (HSNSNS);
 \item[(b)] Realistically spinning neutron star--neutron star systems (RSNSNS);
 \item[(c)] Highly spinning black hole--neutron star systems (HSBHNS),
 \item[(d)] Realistically spinning black hole--neutron star systems (RSBHNS),
 \item[(e)] Spinning black hole--black hole systems (BHBH).
\end{enumerate}
These physical systems are defined by the minimum and maximum values of the 
masses and spins, which we collect in Table~\ref{mass-spin-values}.
\begin{table}[!ht]
\begin{center}
\begin{tabular}{|c|c|c|c|c|c|c|c|c|} \hline
 Type & $m_1^{\text{min}}$ & $m_1^{\text{max}}$ & $m_2^{\text{min}}$ & 
$m_2^{\text{max}}$ & $\chi_1^{\text{min}}$ & $\chi_1^{\text{max}}$ & 
$\chi_2^{\text{min}}$ & 
$\chi_2^{\text{max}}$ \\\hline
HSNSNS & 1 & 2.5 & 1 & 2.5 & 0 & 1 & 0 & 1 \\\hline
RSNSNS & 1 & 2.5 & 1 & 2.5 & 0 & 0.1 & 0 & 0.1 \\\hline
HSBHNS & 5 & 20 & 1 & 2.5 & 0 & 1 & 0 & 1 \\\hline
RSBHNS & 5 & 20 & 1 & 2.5 & 0 & 1 & 0 & 0.1 \\\hline
BHBH & 5 & 20 & 5 & 20 & 0 & 1 & 0 & 1 \\\hline
\end{tabular}
\caption{\label{mass-spin-values}Minimum and maximum values of the masses 
in solar masses and 
spin magnitudes for each system type in our simulations.}
\end{center}
\end{table}

For each system type, we randomized over $10^{4}$ systems with the following 
distributions:
\begin{itemize}
\item the masses $m_i$ are uniformly distributed in log space between a 
maximum and a minimum value;
\item the spin magnitudes $\chi_i$ are uniformly distributed between a 
maximum and a minimum value;
\item all angles are uniformly distributed on the sphere.
\end{itemize}
We did not randomize over distance, as it is factored out in the faithfulness.

\subsection{Results}

Let us first discuss the computational efficiency of our new model. 
This is  difficult because it is highly dependent on 
technical factors, 
such as the duration of the time signal, the Nyquist frequency of the 
time-domain waveform, the discretization of the frequency series, the choice of 
$k\sub{max}$, etc. 
The efficiency is also significantly dependent on the physical parameters of 
the 
system considered, 
such as the spin magnitudes, the spin orientation, and the mass ratio. To give 
a feeling of the computational efficiency of our waveform, we have collected in 
Table~\ref{efficiency} the average 
computation time for our waveforms and for the corresponding time-domain 
waveform both including the time necessary to solve the evolution 
equations, computed from a simulation of 1000 systems of each type described 
above on a modern computer. We have chosen to start our waveform generation at 
10~Hz and stop it at the Schwarzschild innermost stable circular orbit (ISCO), 
i.e. when the PN parameter $\xi$ 
defined in Eq.~\eqref{orbtime} reaches $6^{-1/2}$. We used a frequency 
resolution of 0.1~Hz for the SUA waveforms and a Nyquist 
frequency of 3 times the orbital frequency at the Schwarzschild ISCO for the 
time-domain waveforms. We did not use any filter on the time series for these 
efficiency comparisons, and used for each waveform presented the code 
available in \texttt{lalsimulation}. Note that time-domain waveforms are 
faster to 
compute for black hole--black hole binaries than for neutron star--neutron star 
ones, so much
so that they become prohibitively expensive for the latter, which is precisely 
the system type for which the inspiral part is the most important. On 
the other hand, the SUA waveforms remain relatively cheap for 
all systems, with an up to 2 orders of magnitude improvement
in computational expense relative to time-domain waveforms for neutron star 
binaries. However, for high-mass systems the efficiency gain is small 
to none, showing the limitations of our model.
\begin{table*}[!ht]
\begin{center}
\begin{tabular}{|c|c|c|c|c|} \hline
 System & TD & $k\sub{max} = 0$ & $k\sub{max} = 3$ & $k\sub{max} = 
10$ \\\hline
HSNSNS & 2.08 s & 38.4 ms & 68.1 ms & 104 ms \\\hline
RSNSNS & 1.96 s & 29.2 ms & 38.1 ms & 52.7 ms \\\hline
HSBHNS & 146 ms & 18.4 ms & 23.3 ms & 33.4 ms \\\hline
RSBHNS & 149 ms & 15.8 ms & 21 ms & 30.2 ms \\\hline
BHBH & 25 ms & 8.58 ms & 11.6 ms & 17 ms \\\hline
\end{tabular}
\qquad \qquad 
\begin{tabular}{|c|c|c|c|c|} \hline
 System & TD & $k\sub{max} = 0$ & $k\sub{max} = 3$ & $k\sub{max} = 
10$ \\\hline
HSNSNS & 2.48 s & 58.2 ms & 148 ms & 333 ms \\\hline
RSNSNS & 2.72 s & 49.3 ms & 103 ms & 225 ms \\\hline
HSBHNS & 187 ms & 26.7 ms & 69 ms & 125 ms \\\hline
RSBHNS & 226 ms & 24.9 ms & 54.2 ms & 112 ms \\\hline
BHBH & 27.1 ms & 13.6 ms & 27.6 ms & 54.3 ms \\\hline
\end{tabular}
\caption{\label{efficiency}Average computation times for the 
generation 
of a (left) restricted and (right) full waveform of each system type. TD 
corresponds to the time-domain 
\texttt{SpinTaylorT4} waveform, and the other three are SUA 
\texttt{SpinTaylorT4Fourier} waveforms. The Nyquist frequency of the time-domain 
waveforms is three times the ISCO orbital frequency, and the 
resolution of the frequency-domain waveforms is 0.1~Hz.}
\end{center}
\end{table*}

As a reference, we also ran the same efficiency comparisons for nonprecessing 
systems, to be able to compare our model with existing spin-aligned waveforms. 
As an example we compared our code to the \texttt{SpinTaylorF2} model, fixing 
 the spin magnitudes to zero. As long as the spins are aligned the 
effect of the spin magnitudes on the computation time is small for all 
the waveforms we compared here. To test this, we ran the same simulations 
with maximal aligned and anti-aligned spins, and found that the 
\texttt{SpinTaylorT4} waveform was $\sim 5\%$ faster for anti-aligned spins 
than for zero spins, while it was $\sim 7\%$ slower for aligned spins. We found 
the same trend for the \texttt{SpinTaylorT4Fourier} waveform with
differences of $\sim 8\%$ in both directions. The \texttt{SpinTaylorF2} 
waveform was unaffected. We collect the results for non-spinning binaries in 
Table~\ref{alignedefficiency}. Note that we included the computation times for 
$k\sub{max} > 0$ to give a more complete description of the computational 
efficiency of our model, even though for nonprecessing systems the SUA 
waveform is independent on $k\sub{max}$. Notice that the computation times are 
much smaller than in Table~\ref{efficiency}. The reason for that is that 
when the spins are aligned or zero, the precession time scale disappears from 
the problem, and the relevant time scale for the numerical integration of the 
equations of motion becomes the radiation reaction time scale, which is much 
longer. This renders not only the solving of the equation of motion faster, but 
also the construction and the evaluation of the necessary interpolation 
functions constructed from the solution. Note that for neutron star--neutron 
star 
systems, the $k\sub{max}=0$ \texttt{SpinTaylorT4Fourier} waveforms are slower 
than the \texttt{SpinTaylorF2} waveforms by only a factor 3--9, depending on 
whether we include subdominant harmonics or not. For higher-mass systems which 
are not inspiral dominated, this ratio can go up to a factor of about $20$.
\begin{table*}[!ht]
\begin{center}
\begin{tabular}{|c|c|c|c|c|c|} \hline
 System & TD & F2 & $k\sub{max} = 0$ & $k\sub{max} = 3$ & $k\sub{max} = 
10$ \\\hline
NSNS & 2.18 s & 2.11 ms & 7.57 ms & 11.3 ms & 19.1 ms \\\hline
BHNS & 118 ms & 0.646 ms & 4.65 ms & 5.12 ms & 6.44 ms \\\hline
BHBH & 18.9 ms & 0.326 ms & 4.25 ms & 4.69 ms & 5.65 ms \\\hline
\end{tabular}
\qquad \qquad 
\begin{tabular}{|c|c|c|c|c|c|} \hline
 System & TD & F2 & $k\sub{max} = 0$ & $k\sub{max} = 3$ & $k\sub{max} = 
10$ \\\hline
NSNS & 2.58 s & 2.01 ms & 16.5 ms & 35.6 ms & 71.7 ms \\\hline
BHNS & 196 ms & 0.591 ms & 6.81 ms & 10.3 ms & 18.3 ms \\\hline
BHBH & 25.5 ms & 0.328 ms & 6.78 ms & 10.2 ms & 17.1 ms \\\hline
\end{tabular}
\caption{\label{alignedefficiency}Average computation times for the 
generation 
of a (left) restricted and (right) full waveform of each system type with zero 
spins. TD 
corresponds to the time-domain 
\texttt{SpinTaylorT4} waveform, F2 corresponds to the spin-aligned 
\texttt{SpinTaylorF2} waveform, and the other three are SUA 
\texttt{SpinTaylorT4Fourier} waveforms. The Nyquist frequency of the 
time-domain 
waveforms is three times the ISCO orbital frequency, and the 
resolution of the frequency-domain waveforms is 0.1~Hz.}
\end{center}
\end{table*}

Let us now discuss the faithfulness between the different models listed in the 
previous subsection and the time-domain \texttt{SpinTaylorT4} waveforms.
 Table~\ref{newfaith} shows
the median, lower 1-$\sigma$ and upper 1-$\sigma$ quantiles\footnote{The 
lower and upper and 1-$\sigma$ quantiles are defined respectively as the 
0.1587 and the 0.8413 quantiles, in analogy with the normal distribution.} of 
the distributions of the faithfulness 
between the latter waveforms and the SUA model, the 
DP~\cite{Chatziioannou:2013} 
and the SP models~\cite{Hannam:2013}. In all calculations, the physical 
parameters are kept fixed, 
as described in the previous subsection. Overall, we find that the SUA method
offers unprecedented levels of accuracy, and can reproduce the 
results of 
much more costly 
time-domain models with mismatches (or unfaithfulness $U = 1 - F$) 
smaller than $10^{-4}$, and typically of the order of $10^{-5}$.
\begin{table*}[ht]
\begin{center}
\begin{tabular}{|c|c|c|c|c|c|c|c|c|c|c|c|c|c|c|c|} \hline
  & \multicolumn{3}{c|}{HSNSNS} & \multicolumn{3}{c|}{RSNSNS}& 
\multicolumn{3}{c|}{HSBHNS}& \multicolumn{3}{c|}{RSBHNS}& 
\multicolumn{3}{c|}{BHBH}  \\\hline
 & 16\% & 50\% & 84\% & 16\% & 50\% & 84\% & 16\% & 50\% & 84\% & 16\% & 50\% & 
84\% & 16\% & 50\% & 84\%  \\\hline
$k\sub{max}=0$, RWF & 2.47 & 2.98 & 3.53 & 2.77 & 3.36 & 3.84 & 1.33 & 1.61 & 
1.94 & 1.34 & 1.62 & 1.96 & 1.77 & 2.29 & 2.88 \\\hline
$k\sub{max}=3$, RWF & 2.79 & 3.37 & 3.86 & 2.78 & 3.39 & 3.88 & 1.38 & 1.65 & 
1.97 & 1.38 & 1.65 & 1.98 & 1.83 & 2.44 & 3.11 \\\hline
$k\sub{max}=10$, RWF & 2.79 & 3.37 & 3.86 & 2.78 & 3.39 & 3.88 & 1.38 & 1.65 & 
1.97 & 1.38 & 1.65 & 1.98 & 1.83 & 2.44 & 3.11 \\\hline
$k\sub{max}=0$, FWF & \textbf{2.72} & 3.48 & 4.22 & \textbf{4.5} & 4.92 & 5.32 
& \textbf{2.31} & 3.11 & 
3.92 & \textbf{2.36} & 3.29 & 4.23 & \textbf{2.57} & 3.18 & 3.56 \\\hline
$k\sub{max}=3$, FWF & \textbf{5.16} & 5.8 & 6.27 & \textbf{5.96} & 6.29 & 6.62 
& \textbf{4.16} & 5.09 & 
5.71 & \textbf{4.29} & 5.42 & 5.94 & \textbf{4.46} & 4.9 & 5.13 \\\hline
$k\sub{max}=10$, FWF & \textbf{5.29} & 5.97 & 6.43 & \textbf{6.18} & 6.47 & 6.75 
& \textbf{4.5} & 5.35 & 5.95 & \textbf{4.42} & 5.51 & 6.1 & \textbf{4.51} & 5.03 
& 5.3 \\\hline
Small spins, RWF & 0.916 & 1.63 & 2.36 & 2.5 & 2.75 & 2.94 & 0.236 & 0.533 & 
1.22 & 0.24 & 0.547 & 1.25 & 0.649 & 1.37 & 2 \\\hline
Small spins, FWF & 0.913 & 1.62 & 2.34 & 2.28 & 2.83 & 3.13 & 0.222 & 0.517 & 
1.22 & 0.226 & 0.526 & 1.22 & 0.648 & 1.35 & 2.11 \\\hline
Single spin, RWF & 0.765 & 1.44 & 2.28 & 2.48 & 2.88 & 3.11 & 0.263 & 0.651 & 
1.17 & 0.258 & 0.681 & 1.4 & 0.931 & 1.52 & 2.09 \\\hline
\end{tabular}
\caption{\label{newfaith}
Lower 1-$\sigma$ (16\%), median (50\%), and upper 1-$\sigma$ (84\%) quantiles 
of 
the faithfulness distributions between the time-domain \texttt{SpinTaylorT4} 
waveforms 
and 
(i) the SUA waveforms with a \texttt{SpinTaylorT4} model for the 
evolution of 
the orbital angular momentum and phase with different values of $k\sub{max}$ 
and 
either with the RWF or the FWF model (first 6 rows);
(ii) the small-spin DP model with either the RWF or the FWF model (seventh and 
eighth rows);
and (iii) the single-spin SP model with the RWF model (ninth row). 
All numbers quoted are $-\log_{10}(1 - \rm{F})$, where $\rm{F}$ is the 
faithfulness, e.g.~a value of 4 corresponds to $\rm{F} = 1 - 10^{-4} = 0.9999$.
We put the 16\% quantiles of the FWF SUA results in boldface, as they 
provide a worst-case scenario estimate for the faithfulness of our model.}
\end{center}
\end{table*}

Let us discuss Table~\ref{newfaith} in more detail. 
First, focus on the first three rows only and observe that the faithfulness 
distributions 
for restricted waveform models do not change much between $k\sub{max} > 0$ and 
$k\sub{max} = 0$. 
This implies that the inaccuracies coming from restricting the sum over $k$ in 
Eq.~\eqref{finalWF} 
to only $k = 0$ are comparable or smaller than the inaccuracies coming from 
neglecting 
PN amplitude corrections. 

Second, now concentrating on the first six rows,
observe that the faithfulnesses for the restricted waveform models are 
worse for black hole--neutron star systems than for other system types, 
while this is not true for the full waveform models. This is because 
the amplitude of the subdominant harmonics appearing at next-to-leading PN 
order, $n=1$ and $n=3$, have an overall $(m_1 - m_2)/M$ 
factor. These corrections are suppressed for systems with similar masses, but 
are important for black hole--neutron star systems, which typically have higher 
mass ratios. 

Third, comparing the fourth to sixth rows, observe that the faithfulnesses for 
the full 
waveform models with $k\sub{max} = 0$ are significantly smaller than those for 
the 
same models with $k\sub{max} = 3$. This is much less so when one compares the
$k\sub{max} = 3$ results to the $k\sub{max} = 10$ ones. Analyzing the data 
from our simulations for all $0 \leq k\sub{max} \leq 10$, we find that the 
faithfulness increase with increasing $k\sub{max}$ starts slowing significantly 
at about $k\sub{max} 
= 2$ or $3$. This means that the 
inaccuracies coming from restricting the sum over $k$ in Eq.~\eqref{finalWF} to 
$k = k\sub{max}$ dominate over other sources of error for $k\sub{max} \leq 2$ 
or $3$.

Fourth, let us finally compare the first six rows to the last three and 
observe that the faithfulness distributions are comparable for the DP and SP 
models. 
This implies that the main source of inaccuracy is probably here in 
the discrepancy 
between 
the \texttt{TaylorT4} phase and the \texttt{TaylorF2} phase, which is common to 
the DP and SP
models, and has been shown to be 
substantial~\cite{buonanno2009,Nitz2013}.
Unlike for the SUA model, including subdominant 
harmonics 
does not here appear to yield a significant increase in faithfulness. 
Furthermore, the only system type for which these models offer 
sufficiently high faithfulnesses is the realistically spinning neutron 
star--neutron star 
type, for which both spin magnitudes are smaller than 0.1 and precession 
effects are thus suppressed. 

\begin{figure*}[ht]
 \includegraphics[width=0.95\columnwidth]{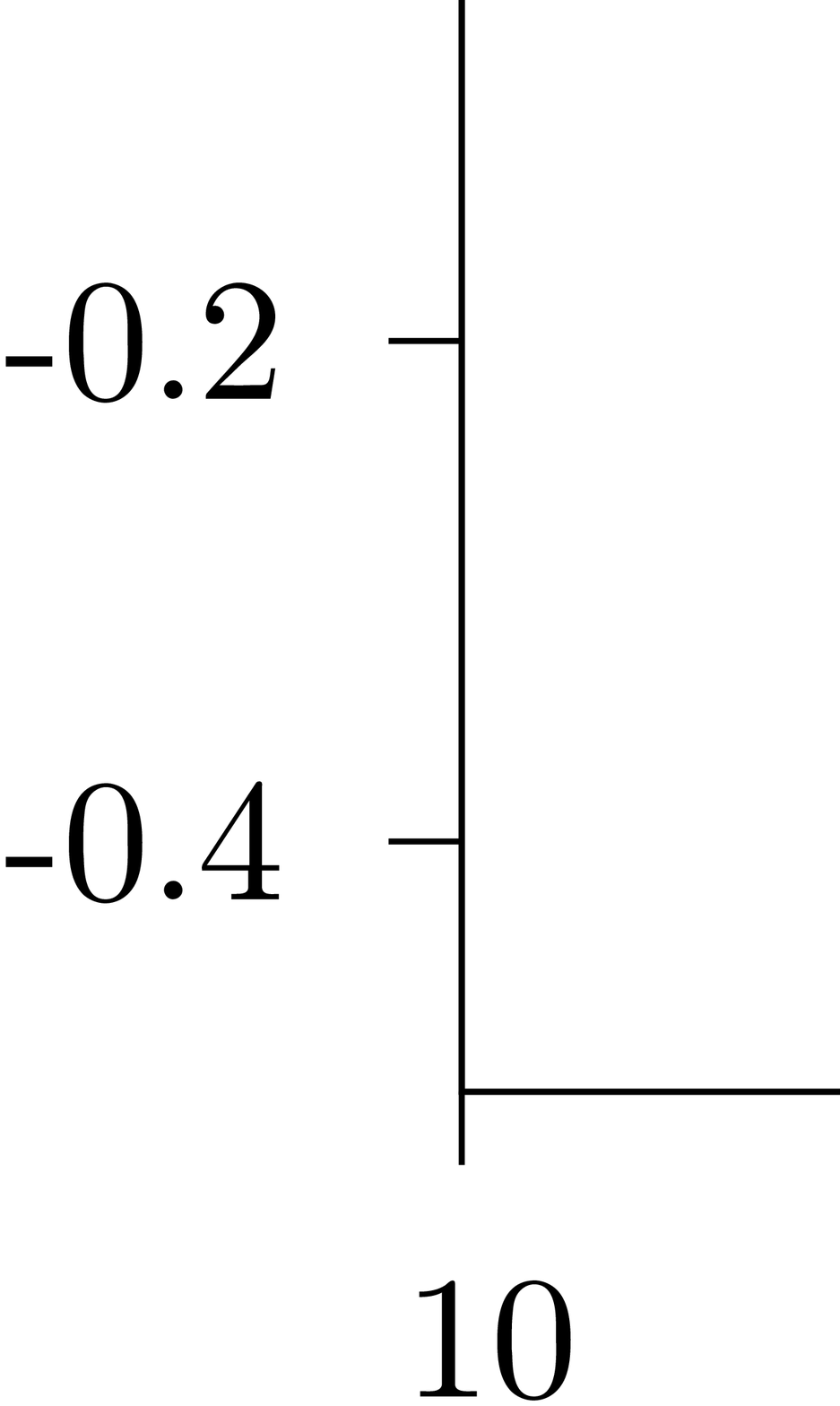}
 \includegraphics[width=0.95\columnwidth]{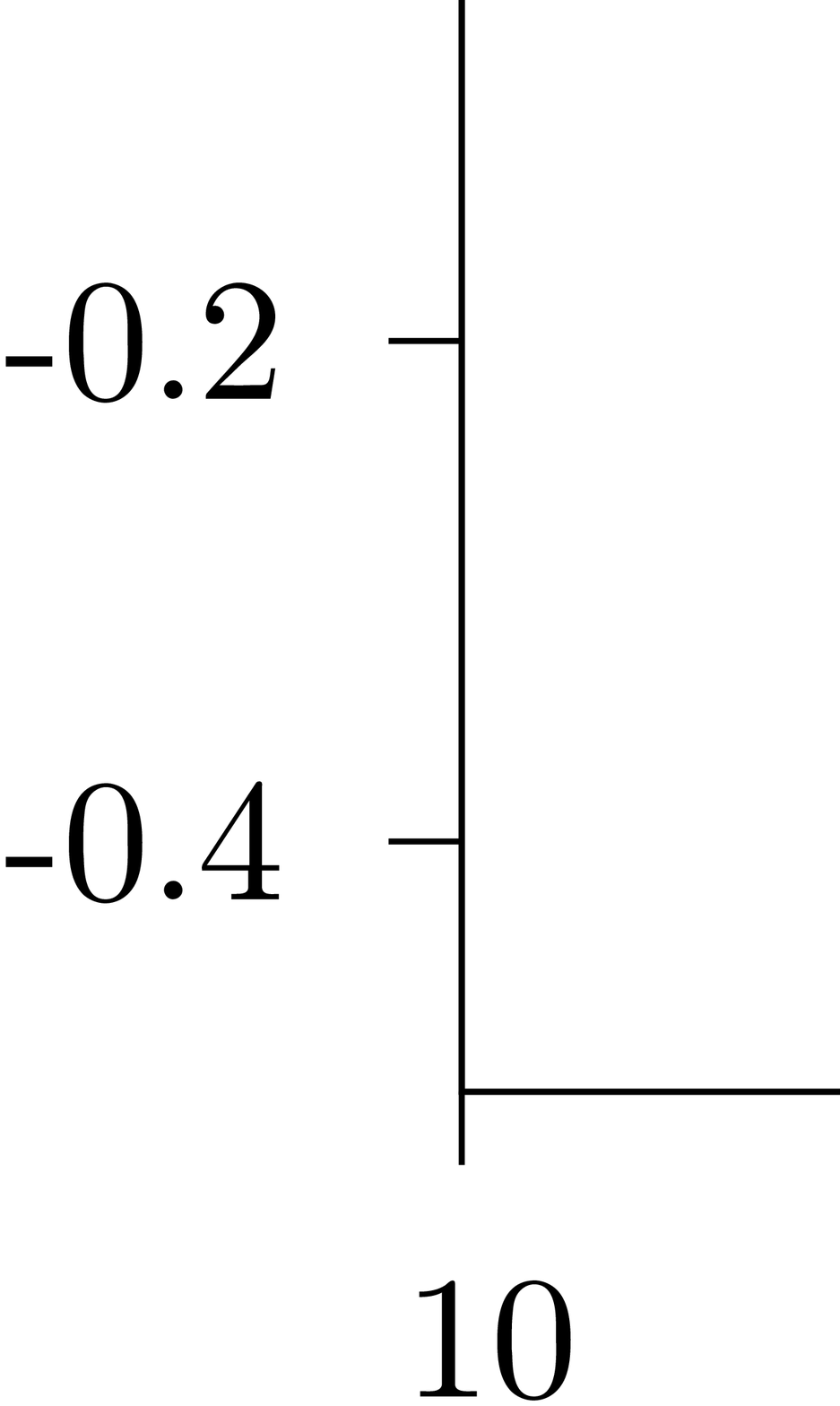}
 \caption{\label{dephasing}Dephasing in radians as a function of GW frequency 
between the time-domain \texttt{SpinTaylorT4} waveforms and the SUA waveforms 
(RWF on the left and FWF on the right) with different  values of $k\sub{max}$ 
($k\sub{max} = 0$ in dashed green, $k\sub{max} = 3$ in dotted red, and 
$k\sub{max} = 10$ in solid blue). 
This figure uses the HSNSNS system with $(m_1,m_{2}) = (2,1.4) M_\odot$ 
and $\chi_1 = 0.9 = \chi_2$ with random orientations.  The unfaithfulness $U = 
1 
- F$ 
for these systems is $10^{-2.49}$, $10^{-2.93}$, and $10^{-2.93}$ for 
$k\sub{max}=0$, 3, and 10 
respectively all with the RWF model, while it is equal to $10^{-2.67}$, 
$10^{-4.92}$, and $10^{-5.19}$ 
for the FWF model.}
\end{figure*}

Let us now try to determine in more detail the reason why the faithfulness is 
so 
much 
better when using the SUA model. Figure~\ref{dephasing} shows 
the 
dephasings, i.e.~the Fourier phase difference, between the 
time-domain \texttt{SpinTaylorT4} waveforms and the SUA 
\texttt{SpinTaylorT4} model (RWF on the left and FWF 
on the right)
with different values of $k\sub{max}$. This comparison uses the HSNSNS system 
type, 
with $(m_1,m_{2}) = (2,1.4) M_\odot$ and $\chi_1 = 0.9 = \chi_2$, with 
orientations chosen randomly. 
Observe that the dephasing for high values of $k\sub{max}$, i.e.~$k\sub{max}=3$ 
or $10$, 
is dominated by an effect appearing on the orbital time scale when using the 
RWF 
model. 
This is because the different harmonics present in the 
time-domain waveform 
differ in frequency by multiples of the orbital frequency, which creates 
beating 
on the orbital scale.
Since only the leading PN order harmonic is present in the RWF model (left 
panel), this 
beating is absent in them, creating a phase discrepancy on the orbital scale. 
Observe, however, 
that this discrepancy disappears when we include subdominant harmonics in the 
SUA waveforms (right panel). 
Observe also that in neither panel does there appear to be a secular growth in 
the 
dephasing, although the amplitudes of the oscillatory dephasings appear 
to be growing with frequency. The phase
discrepancy coming from restricting the sum over $k$ in Eq.~\eqref{finalWF} is 
visible in the $k\sub{max} = 0$ curve, and is comparable in amplitude to the 
discrepancy coming from neglecting subdominant harmonics, visible in the higher 
$k\sub{max}$ RWF curves.

\section{Conclusion}
\label{conclusion}

We have presented here a new shifted uniform asymptotic method for 
constructing the Fourier transform of the GW response function of 
an inspiraling binary system. This method requires as input only the time 
evolution of the orbital 
angular momentum and of the orbital phase and frequency, which can be provided 
either 
analytically or numerically. The output of the method, the Fourier transform of 
the GW response, is highly accurate when compared to
waveforms obtained through discrete Fourier transforms of time-domain signals, 
even for gravitational 
waves generated in generic spin precessing inspirals. This method avoids the 
catastrophes that lead to singularities in the Fourier amplitude 
obtained with a simple stationary phase approximation. Moreover, the method is 
computationally efficient, allowing for the construction of frequency-domain 
templates 50 times faster than time-domain waveforms.  

We then provided an example of this method by applying the SUA 
method to a particular model for the time evolution of the orbital angular 
momentum, the orbital phase and the frequency, the \texttt{SpinTaylorT4} model. 
We then 
compared the resulting Fourier transform to the corresponding time-domain 
waveforms produced by the discrete Fourier transform of a time signal through 
the faithfulness measure. We found unfaithfulnesses of order 
$10^{-4}$--$10^{-6}$, a 3--4 
orders of magnitude improvement over other frequency-domain templates 
currently 
used in GW data analysis. 

We expect the SUA method to be very useful in GW 
astronomy, as it is a highly computationally efficient and generic method that 
can be applied to any model in the time domain. SUA waveforms 
are faithful to full time-domain waveforms obtained through discrete Fourier 
transforms, and thus, they allow for 
small parameter biases induced by template mismodeling. Moreover, their 
computational efficiency should allow for large-scale parameter estimation 
studies with advanced, ground detectors. 

\acknowledgments
We would like to thank Richard O'Shaughnessy and Katerina Chatziioannou for 
useful comments and suggestions. 
N. Y. acknowledges support from NSF Grant No. PHY-1114374 and the NSF CAREER 
Grant No. 
PHY-1250636, as well as support provided by the National Aeronautics and Space 
Administration from Grant No. NNX11AI49G, under Grant No. 00001944. N. J. C. 
acknowledges 
support from NSF Grant No. PHY-1306702. A. K. is supported by NSF CAREER Grant 
No. 
PHY-1055103.

\appendix
\section{Equations of motion}
\label{app:eqmot}

For completeness, we here give the full expression for the equations of motion 
in the \texttt{TaylorT4} form, 
in terms of the dimensionless individual masses $\mu_1 = m_1/M$, and $\mu_2 
= m_2/M$ and the dimensionless spins $\bm{a}_1 = \bm{S}_1/m_1 M $ and $\bm{a}_2 
= \bm{S}_2/m_2 M $.

The 3.5PN radiation-reaction equation with 2PN spin-spin coupling is given 
by~\cite{racinebuonannokidder,Favata:2009,blanchet3PN,Bohe:2013cla} 
\begin{align}
 M\dot{\xi} &= \xi^9 \sum_{n = 0}^{7} b_n \xi^n, \\
 b_0 &= \frac{32 \nu}{5}, \\
 \frac{b_2}{b_0} &= -\frac{743}{336} - \frac{11}{4}\nu, \\
 \frac{b_3}{b_0} &= 4\pi - \beta_3, \\
 \frac{b_4}{b_0} &= \frac{34103}{18144} + \frac{13661}{2016}\nu + 
\frac{59}{18}\nu^2 - \sigma_4,\\
 \frac{b_5}{b_0} &= -\pi \left(\frac{4159}{672} + \frac{189}{8} \nu \right) - 
\beta_{5},\\
 \frac{b_6}{b_0} &= 
\frac{16447322263}{139708800} - \frac{56198689}{217728}\nu + \frac{541}{
896} \nu^2 - \frac{5605}{2592} \nu^3 \nonumber\\
&+ \pi^2 \left(\frac{16}{3} + 
\frac{451}{48} \nu\right) 
- \frac{1712}{105}\left[ \gamma_E + \log(4\xi) \right] - \beta_6,\\
 \frac{b_7}{b_0} &= 
-\pi\left(\frac{4415}{4032} - \frac{358675}{6048} \nu - \frac{91495}{1512}
 \nu^2 \right)- \beta_{7},
\end{align}
where $\gamma_E$ is the Euler constant, and the spin-orbit couplings $\beta_i$ 
and the spin-spin coupling $\sigma_4$ 
are given by
\begin{align}
\beta_{3} &= \sum_{A\neq 
B} 
\left(\frac{113}{12} \mu_A + 
\frac{25}{4} \mu_B\right) \uvec{L} \cdot \bm{a}_A ,
\\
\beta_{5} &= \sum_{A\neq 
B} \bigg[ \left( \frac{31319}{1008} 
- \frac{1159}{24}\nu \right) \mu_A \nonumber\\
&+ 
\left(\frac{809}{84}-\frac{281}{8}\nu\right) \mu_B \bigg] \uvec{L} \cdot 
\bm{a}_A ,
\\
\beta_6 &= \pi \sum_{A\neq 
B} 
\left(\frac{75}{2} 
\mu_A + \frac{151}{6} \mu_B \right) \uvec{L} \cdot \bm{a}_A ,\\
\beta_{7} &=  \sum_{A\neq 
B}\bigg[ \left( \frac{130325}{756} - \frac{796069}{2016}\nu + \frac{100019}{864}
\nu^2\right) \mu_A \nonumber\\
& + \left(\frac{1195759}{18144} - \frac{257023}{1008}\nu + \frac
{2903}{32}\nu^2 \right) \mu_B \bigg] \uvec{L} \cdot \bm{a}_A ,
\\
\sigma_4 &= \frac{247}{96} \left( \bm{a}_1 + \bm{a}_2 \right)^2 - 
\frac{721}{96} \left[ \uvec{L} \cdot \left( \bm{a}_1 + \bm{a}_2 \right) 
\right]^2 \nonumber\\
&- \sum_A \left[ \frac{7}{48} \bm{a}_A^2 - \frac{1}{48} \left( \uvec{L} \cdot 
\bm{a}_A \right)^2 \right].
\end{align}

The full 2PN spin-spin and 3.5 PN spin-orbit precession equations are given 
by~\cite{PhysRevD.12.329,Bohe:2012mr}
\begin{align}
 M\duvec{L} &= - \xi^6 \left( \bm{\Omega}_1 + 
\bm{\Omega}_2 \right),  \\
 M\dvec{a}_1 &= \mu_2 \xi^5 \bm{\Omega}_1,\\
 M\dvec{a}_2 &= \mu_1 \xi^5 \bm{\Omega}_2,\\
 \bm{\Omega}_A &= \left( C_{A,0} +  C_{A,2} \xi^2 + C_{A,4} \xi^4 + D_A \xi 
\right) 
\uvec{L} \times \bm{a}_A \nonumber\\
&+ \frac{1}{2} \xi \, \bm{a}_B \times \bm{a}_A, \\
C_{A,0} &=  2 \mu_A + 
\frac{3}{2} 
\mu_B, 
\\
C_{A,2} &=  3 \mu_A^3 + 
\frac{35}{6} \mu_A^2 \mu_B + 4 \mu_A \mu_B^2 + \frac{9}{8} \mu_B^3, 
\\
C_{A,4} &=  \frac{27}{4} 
\mu_A^5 + \frac{31}{2} \mu_A^4 \mu_B + \frac{137}{12} \mu_A^3 \mu_B^2 
\nonumber\\
&+ 
\frac{19}{4} 
\mu_A^2 \mu_B^3 + \frac{15}{4} \mu_A \mu_B^4 + \frac{27}{16} \mu_B^5 , \\
D_A &= - \frac{3}{2} \uvec{L} \cdot \bm{a}_B,
\end{align}
where it is understood that $A,B \in \{ 1,2\}$ and $A \neq B$.

\section{Alternative derivation of the SUA constants}
\label{app:stencils}

A 
$(2k\sub{max}+1)$-point stencil of the point $t_0$ is the collection of points
\begin{align}
S_{k\sub{max}} &= \{t_0 - k\sub{max} T, t_0 - 
(k\sub{max}-1) T, \ldots, t_0 + k\sub{max} T \}. 
\end{align}

$S_{k\sub{max}}$ can be used to approximate the first $2k\sub{max}+1$ 
derivatives of a function $f(t)$ at $t=t_0$ (including the zeroth order 
derivative):
\begin{align}
 T^m f^{(m)} (t_0) &\approx \sum_{k = - k\sub{max}}^{k\sub{max}} 
b_{k,m,k\sub{max}} 
f(t_0 + k T).
\end{align}

To compute the coefficients $b_{k,m,k\sub{max}}$, one needs to expand the 
functions $f(t_0 + k T)$ on the right-hand side of the equation above as a 
power 
series in $T$ up to order $2k\sub{max}$. Solving the equation order by order in 
$T$, we get a unique solution for the $b_{k,m,k\sub{max}}$. We can also extend 
this procedure to arbitrary high order derivatives, and we get 
$b_{k,m,k\sub{max}} = 0$ when $m > 2k\sub{max}$. Symmetry ensures that  
$b_{k,m,k\sub{max}} 
= (-1)^m b_{-k,m,k\sub{max}}$. The system of equations defining the constants 
$b_{k,m,k\sub{max}}$ is
\begin{align}
 \sum_{k=-k\sub{max}}^{k\sub{max}} \frac{k^n}{n!} b_{k,m,k\sub{max}} = 
\delta_{n,m}.
\end{align}

With this in hand, we start from Eq.~\eqref{Hcorr_intermediate}, and 
approximate each time derivative using the stencil $S_{k\sub{max}}$:
\begin{align}
 \tilde{H}\sub{corr}(f) &\approx \sum_{p \geq 0} \frac{\left(-i\right)^p}{2^p 
p!}  
\sum_{k=-k\sub{max}}^{k\sub{max}} b_{k,2p,k\sub{max}} e^{- i \delta\phi(t_0 + k 
T)} \\
&= \sum_{k=0}^{k\sub{max}} \sum_{p = 0}^{k\sub{max}} 
\frac{\left(-i\right)^p}{2^p 
p!}   b_{k,2p,k\sub{max}}\nonumber\\
&\times\left[ e^{- i 
\delta\phi(t_0 + k 
T)} + e^{- i \delta\phi(t_0 - k 
T)} \right]
, \label{stencilresult}
\end{align}
where in the last equality we used the facts that all $b_{k,2p,k\sub{max}}$ 
vanish when $p > k\sub{max}$ and that all derivatives are of even order, and we 
reordered the sums. 
We can rewrite Eq.~\eqref{stencilresult} in a form similar to 
Eq.~\eqref{Hcorr2}, 
with the coefficients $c_{k,k\sub{max}}$ taking the place of the 
$a_{k,k\sub{max}}$, and defined by
\begin{align}
 c_{k,k\sub{max}} &= 2\sum_{p = 0}^{k\sub{max}} 
\frac{\left(-i\right)^p}{2^p 
p!}   b_{k,2p,k\sub{max}}.
\end{align}

\pagebreak

We can
multiply by $k^{2n}/(2n)!$, and sum over $k$. We get
\begin{align}
 \sum_{k=0}^{k\sub{max}} c_{k,k\sub{max}} \frac{k^{2n}}{(2n)!} &= 
\sum_{p=0}^{k\sub{max}}\frac{\left(-i\right)^p}{2^p 
p!} \times 2 \sum_{k=0}^{k\sub{max}}  \frac{k^{2n}}{(2n)!}  b_{k,2p,k\sub{max}} 
\\
&=
\sum_{p=0}^{k\sub{max}} \frac{\left(-i\right)^p}{2^p 
p!} \delta_{n,p} = \frac{\left(-i\right)^n}{2^n 
n!}
\end{align}

This system of equations accepting only one solution, we have $c_{k,k\sub{max}} 
= a_{k,k\sub{max}}$ for any $\{k,k\sub{max}\}$.

\bibliography{master}
\end{document}